# H.264/SVC Mode Decision Based on Mode Correlation and Desired Mode List

**L. Balaji**
Department of ECE, Velammal Institute of Technology, Chennai, India
E-mail: maildhanabal@gmail.com

**K. K. Thyagharajan**
ECE Department, RMD Engineering College, Chennai, India
E-mail: kkthyagharajan@yahoo.com

**Abstract:** The design of video encoders involves the implementation of fast mode decision (FMD) algorithm to reduce computation complexity while maintaining the performance of the coding. Although H.264/scalable video coding (SVC) achieves high scalability and coding efficiency, it also has high complexity in implementing its exhaustive computation. In this paper, a novel algorithm is proposed to reduce the redundant candidate modes by making use of the correlation among layers. The desired mode list is created based on the probability to be the best mode for each block in the base layer and a candidate mode selection in the enhancement layer by the correlations of modes among the reference frame and current frame. Our algorithm is implemented in joint scalable video model (JSVM) 9.19.15 reference software and the performance is evaluated based on the average encoding time, peak signal to noise ratio (PSNR) and bit rate. The experimental results show 41.89% improvement in encoding time with minimal loss of 0.02dB in PSNR and 0.05% increase in bit rate.

## 1 Introduction

Applications of multimedia through digital broadcasting over various kinds of devices (like mobile, laptop, personal data assistants (PDAs), high definition television (HDT), standard definition television (SDTV), etc.) are increasingly important. And they need better scalability in video coding due to the variable nature of bandwidth. A scalable extension of H.264/advanced video coding (AVC) is standardized to provide the best suitable video coding in 2007 as H.264 scalable video coding[1]. Reference software is developed by the motion picture experts group (MPEG) and video coding experts group (VCEG) jointly called a joint video team (JVT) for scalable video coding[2,3].

The inherent nature of spatial, temporal and signal to noise ratio (SNR) or quality scalability with respect to H.264/AVC makes H.264/scalable video coding (SVC) standardized[3], and its performance in achieving high efficiency in coding is evaluated[4]. In spatial scalability, the picture with the lowest spatial resolution is considered as base layer and is encoded as H.264/AVC compatible bitstream, whereas the picture with high resolution which is an unsampled residue between the original and reconstructed signal of the base layer is considered the enhancement layer. In temporal scalability, a hierarchical B picture approach is used for a particular spatial layer with zero structural delay. H.264/SVC constitutes I, P and B pictures in which I/P picture will be the key picture and is encoded with normal intervals by the only previous key picture as reference. The B picture encodes the pictures between the two key pictures. The size of the group of pictures (GOP) size determines the number of temporal layers in a spatial layer, where a GOP is nothing but a key picture followed by all the temporally located pictures till the next key picture. The relation between the spatial and temporal scalability employs SNR or quality scalability which is based on different spatio-temporal reconstruction quality levels namely coarse-grain scalability (CGS) and medium grain scalability (MGS). CGS is nothing but a single temporal layer per spatial layer and MGS is multiple temporal layers per spatial layer.

Although H.264/SVC with a unique bitstream adaptation to various bit rates, transmission channel bandwidth and display capabilities, achieves high scalability and high efficiency in coding, the computation complexity of the encoder is very high because of its inherent nature. Due to the hierarchical B picture approach in the temporal layer, it needs all the modes to be searched to be the best candidate mode prediction by full search algorithm implemented in joint scalable video model (JSVM). This is more time consuming and complex for the encoder. Focusing this issue, many research works were proposed to reduce the complexity in terms of fast mode decision (FMD) algorithm by reducing the redundant candidate mode in H.264/SVC. These works predict the redundant modes using rate-distortion cost (RDC) function and the correlation among the hierarchical B picture structure. The computation complexity was efficiently decreased by these works with the degraded video quality. But they were not suitable for sequences with large motions.

Nowadays, too many handheld devices with typical structural implementations, have an increasing requirement for video quality as an important issue. It is an enhancement layer where the quality has to be increased. But to conserve power for handheld devices is also an important issue to be considered particularly for real-time video applications. Overall, the video quality and reduction in computation complexity need to be more important while implementing any algorithm. In this paper, we focus on the reduction of candidate mode by using the probability model and mode correlation. The probability model creates a list of modes to be the best in the base layer and mode correlation decides the best mode in the enhancement layer. The rest of this paper is organized as follows. In Section 2, background and related works based on fast mode decision algorithms implemented in SVC, rate-distortion cost procedure, probabilistic model and mode correlations were discussed. In Section 3, the proposed algorithm for complexity reduction is discussed. And the experimental results with comparative analysis are discussed in Section 4. Section 5 concludes this paper.

## 2 Background and related work

Three new modes such as motion vectors, residuals, and intra information were introduced in the inter-layer prediction from the base layer to select the best coding mode in the enhancement layer. Based on these inter-layer prediction modes, better improvement in coding efficiency is achieved along with scalability. But these inter-layer modes have to do ratedistortion optimization (RDO) many times which involves very high computational complexity.

Particularly, residual prediction mode must be performed twice of the RDO process which increases twice the computational complexity of the normal RDO process of H.264/AVC. This complexity implementation is reduced by an efficient architecture proposed in [5] by changing the processing order, here the prediction mode of reference macroblock (MB) is used to predict the candidate modes. It follows two steps if the prediction mode of reference MB is an INTRA or INTER mode. If it is INTRA mode, the INTRA 4 × 4 mode is checked with 8 × 8 mode. And if 4 × 4 mode is smaller, it will be selected. If 8 × 8 mode is smaller, INTER 8 × 16 mode in the base layer is checked with 16 × 8 mode, the smaller one will be selected. If it is INTER mode, the prediction is based on motion vectors (MV). If MV is greater than a threshold, the best mode is selected among three candidates of upper MB mode, left MB mode and SKIP mode, else the reference MB is checked with the SKIP mode. This approach achieves faster encoding but with degradation in performance. The best mode is selected by checking both the reference MBs[6]. If both are encoded as SKIP modes, SKIP mode is selected, else the mode saved in the previous MB is selected. If any of the reference MB is SKIP mode, then 16×16 mode is checked with SKIP mode, the smaller will be selected. This approach achieves better performance than that in [5] and improves rate-distortion, but the encoding speed is not much faster. In [7], a very high complexity reduction rate is achieved by decreasing some candidate modes in H.264/SVC. However, the previous works are based on sacrificing the video quality of the enhancement layer by reducing the complexity rate. In our previous work[8], an adaptive rate control scheme is proposed which reduces bit rate and maintains PSNR. Here, initial quantization parameter is estimated based on Cauchy probability density function (PDF). This function dynamically adjusts the selection of mode which is slightly better than full search algorithm. Low complexity algorithm for addressing the computation complexity in [9] reduces the candidate modes by mode correlations between the base layer and the enhancement layers. For an MB to be coded either by INTRA or INTER type, two algorithms were proposed which decrease the redundant candidate modes, but the quality of the enhancement layer degrades. In [10], INTER modes are divided into a set of SKIP mode and a set of non-SKIP mode. All non-SKIP modes will have almost the same RDC except SKIP mode. Here, the SKIP mode is checked initially with the reference MB. If the reference MB to be encoded is not SKIP mode, INTER 16 × 16 and 8 × 8 is checked to derive various fast mode decision techniques. The non-SKIP mode is again divided into three groups to select the desired mode based on the threshold and model parameter. This probability model implemented for H.264/AVC achieves fast mode selection and is not implemented in H.264/SVC. An efficient mode reduction method is proposed in [11] to decrease the encoding complexity, but it is not suitable for the sequences with large motions.

## 2.1 Rate distortion cost

RDO used in JSVM is used to estimate the RDC for all modes using the Lagrangian parameter ($λ$). RDC is a cost function of rate and distortion with $λ$. This $λ$ acts as a weighting parameter which adjusts the bit cost as

$$J(s,c,MODE|QP,λ_{MODE}) = SSD(s,c,mode|QP) + λ_{MODE} R(s,c,MODE|QP) \quad (1)$$

where $λ_{MODE}$ is the weighting parameter or Lagrangian parameter, $s$ is the original MB, $c$ is the reconstructed MB, $SSD$ is the sum of squared differences between $s$ and $c$, $QP$ is the quantization parameter, and $R$ is the number of bit cost for motion vector, header information and the discrete cosine transform (DCT) coefficients.

## 2.2 Probability model

A desired mode list based on the probability of the mode to be the best possible mode with the co-located macro block has been evaluated in [10]. There is a correlation between an MB to be coded and its neighboring MB. With this correlation, a desired mode list is created with the highest probability to be the best mode. Therefore, a reference MB set *P* is described as

$$P = \begin{Bmatrix} MB(j, x-16, y-16), MB(j, x-16, y) \\ MB(j, x, y-16) \\ MB(j, x+16, y-16), MB((j-1), x, y) \end{Bmatrix} \quad (2)$$

where *MB(j,x,y)* denotes the MB located at the *j*-th frame with upper left pixel *x,y*, and $MB((j-1), x, y)$ denotes the previous co-located MB located at (*j*−1)-th frame (the same as the current coding MB) with upper left pixel *x,y*. The neighboring MB mode set *Q* is given by

$$Q = \{M_{MB} | MB \in P\} \quad (3)$$

where $M_{MB}$ denotes the encoding mode of MB. The approximated probability[9] of the mode to the best mode is given by the probability model as

$$P(m = M | M \in Q) \approx P(m \in P) \frac{\sum_{MB \in P, M_{MB}=m} N(m = M_{MB})}{\sum_{(m \in Q)} \sum_{MB \in P, M_{MB}=m} N(m = M_{MB})} = K \times \sum_{MB \in P, M_{MB}=M} N(m = M_{MB}) \quad (4)$$

where *N*(·) is the occurrence time of an event and *K* is constant. Here, $M_{MB}$ is not an element of *Q*, which has less probability to be the best mode and is considered to be zero.

## 2.3 Mode correlation

There is a high degree of correlation for a selected candidate mode between a reference frame and the best mode of current frame[11], which is evaluated as follows a 4 layer model,

1) When both reference 0 frame and reference 1 frame are SKIP modes, best mode is a SKIP and 16 × 16.

2) When one of the reference frames is a SKIP mode, best mode is SKIP and 16 × 16 or other modes.

3) When none of reference frames is in SKIP mode, the best mode is surrounding modes of reference MB, also with SKIP and 16 × 16.

4) Reference MBs along with SKIP and 16 × 16 is the best mode for other layers.

Also, the correlation among the redundant candidate modes with reference frames increases for higher layers.

## 3 Proposed algorithm

The objective of this work is to reduce the encoding time while maintaining almost the same PSNR level and bit rate, by selecting the desired candidate mode from all modes. We start up with two strategies in our algorithm for base and enhancement layers. If the co-located MB is inter coded in the base layer, we make use of mode correlation among the temporal layers on selection of a desired candidate mode, otherwise all the intra modes are checked for the best mode.If the co-located MB is inter coded in the enhancement layer, we make use of the correlations between modes. As moving on to the higher layers, the motion vector difference (MVD) becomes lesser. So we decide upon checking upper, left, right and bottom modes of the current MB. If this selection is not best suited, (4) will decide upon selecting the best nonSKIP mode. This selection of non-SKIP mode works on three levels, the first selection is based on checking if it is 16 × 16 or SKIP mode. If it is 16 × 16, we go on checking the lower order 16 × 8, 8 × 16, and all sub blocks of 8 × 8 as shown in the algorithm.

The RDC of the best current MB mode selected is compared with the minimum RDC among all the modes as given in (5), then the current mode will be taken as the best mode, and all the other modes are omitted.

$$J{\text{curr}} < J{\text{min}} \qquad (5)$$

where $J_{\text{curr}}$ denotes RDC of the current MB mode and $J_{\text{min}}$ denotes the minimum RDC of all the modes obtained so far.

The threshold level ($J_{\text{TH}}$) for each enhancement layer is obtained as a product of minimum RDC among all modes with a model parameter Γ given in (6). The values for the model parameter Γ shown below for each layer is obtained after a series of simulation as $J\text{TH} = Γ × J\text{min}$

$$\Gamma = \begin{cases} 0.6, & \text{Enhancement layer 1} \\ 0.9, & \text{Enhancement layer 2} \\ 1.2, & \text{Enhancement layer 3}. \end{cases} \qquad (6)$$

This threshold ($J_{\text{TH}}$) will be the minimum RDC($J_{\text{min}}$) for each enhancement layer.

**Algorithm 1.**
  **Step 1.** Start encoding a macro block in base layer;
   If (co-located MB==inter coded)
    Calculate JSKIP and J16×16 modes; If (Ref frame 0 || Ref frame 1==SKIP)
      If (Ref frame 0 && Ref frame 1==SKIP) Best candidate mode is from SKIP and
     16 × 16 using (1)
      and (5);
    Else
      Best candidate mode is decided for the other modes such as 16 × 8, 8 × 16 and
     8 × 8 using (1) and (5);
    Else
     Go to Step 2;
   Else
    Check all the intra modes for the best mode.
  **Step 2.** If (co-located MB==inter coded Enh layer1)

If (co-located MB==inter coded Enh layer2) Calculate RDC (J) for surrounding MB of reference frame 0 and reference frame 1 such as upper, left, right and bottom and check for the best candidate mode using (1), (5) and (6); Go to Step 1;
  Else
    Calculate RDC (J) for the reference frames 0 and 1, and check for the best candidate mode using (1), (5) and (6); Go to Step 1;
  Else
    Go to Step 3.

**Step 3.**
Construct a desired mode list with their probabilities to be the best with (4) and add other modes in the order of SKIP, intra 16 × 16, intra 16 × 8, intra 8 × 16 and P 8 × 8 which are not in Q as in (3).

**Step 4.**
  If (ModeCurr== P 8 × 8)
    If (Modefirst==SKIP intra 16 × 16) Best candidate mode is decided from 8 × 8, and intra 8 × 8 using (1), (5) and (6);
    Go to Step 1;
    Else If (Modefirst==intra 16 × 8) Best candidate mode is decided from DIRECT8 × 8, intra 8 × 8 & intra 8 × 4 using (1), (5) and (6);
    Go to Step 1;
    Else If (Modefirst==intra 8 × 16) Best candidate mode is decided from DIRECT8 × 8, intra 8 × 8 and intra 4 × 8 using (1), (5) and (6);
    Go to Step 1;
  Else
    Best candidate mode is decided from all the sub blocks using (1), (5) and (6); Go to Step 1;
  Else
    Go to Step 5.

**Step 5.**
  If (ModeCurr ($J$curr) >$J$min)
    Check all the intra modes for the best mode using (1), (5) and (6); Go to
    Step 1; Else Go to
    Step 3.

## 4 Experimental results

The H.264/SVC reference software JSVM 9.19.15 is used for evaluating the results[12]. Low complexity macro block mode is set to evaluate the results for JSVM in the reference software. We use time, PSNR and bit rate to evaluate the performance of the original and test algorithm. Nine video sequences of common intermediate format (CIF) such as Bus, Foreman, Football, Mobile, City, Crew, Ice, Harbour and Soccer are chosen. The layers of each sequence were set with the quantization parameter values 28, 34 and 40. The GOP size is set as 30 frames per second in our algorithm.

$$\Delta time = \frac{time_{proposed} - time_{JSVM}}{time_{JSVM}} \times 100\% \qquad (7)$$

$$\Delta PSNR = PSNR_{proposed} - PSNR_{JSVM} \qquad (8)$$

$$\Delta bitrate = ((bitrate_{JSVM} - bitrate_{proposed}) / bitrate_{JSVM}) \times 100\%. \qquad (9)$$

The simulation results are shown in Table 1. From the results, the average encoding time of the proposed method is 41.89% faster than JSVM with a loss of 0.02dB in PSNR and 0.05 % in terms of bit rate.

Fig.1 shows the average encoding time of each sequence encoded with JSVM and the proposed algorithm. From the observations, the difference in encoding time between our proposed algorithm and JSVM in the foreman sequence is the largest. It is less than half of the encoding time of JSVM with our algorithm. This is due to the correlation among modes between the current frame MB and the reference frame MB. At the same time, the minimum encoding time difference between the proposed algorithm and JSVM is the Harbour sequence. Here, the encoding time of our algorithm is not much faster than JSVM. This is because the sequence involves more spatial and temporal details. For large changes in the sequence, we apply the probability model. It takes a few more time to compute RDC of all the modes in the non-SKIP mode, but results in good PSNR quality with the same bit rate.

So, the probability mode does an exhaustive search for modes to bring better quality of the signal, but the encoding speed is somewhat slower as compared to the correlation among modes between the current and the reference frames.

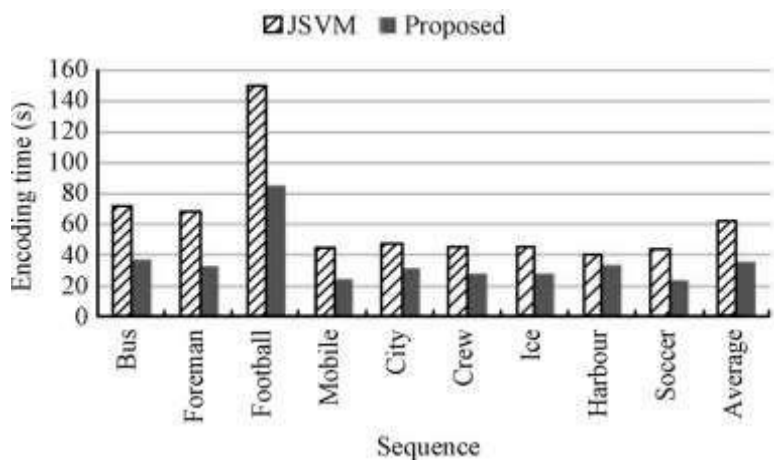

Fig.1 Comparison of the encoding time between JSVM and the proposed algorithm

Fig.2 shows the differences in PSNR between the proposed algorithm and JSVM. And from the observations, the sequences such as Bus, Foreman, Mobile have less PSNR levels as compared to JSVM. All the other sequences have similar PSNR levels. Fig.3 shows the difference in terms of bit rate between the proposed algorithm and JSVM. Reduction in bit rate is achieved between Foreman, Soccer and Bus sequences through our algorithm. Also, an overall bit rate reduction of 0.5% for these three sequences are achieved.

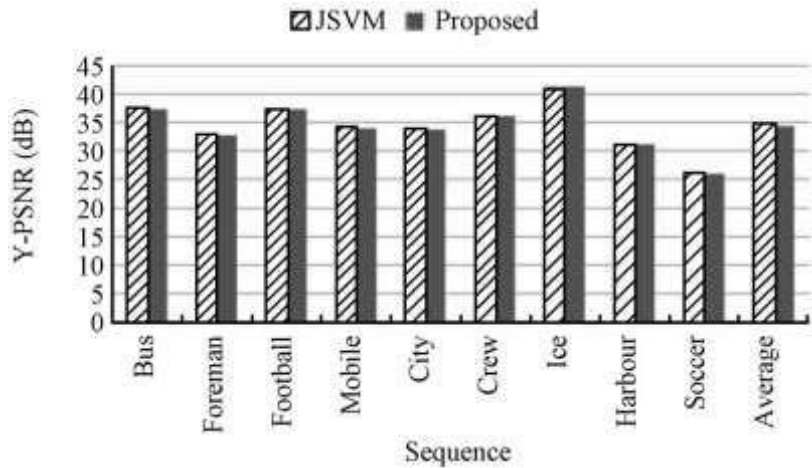

Fig.2 Comparison of the average PSNR between JSVM and the proposed algorithm

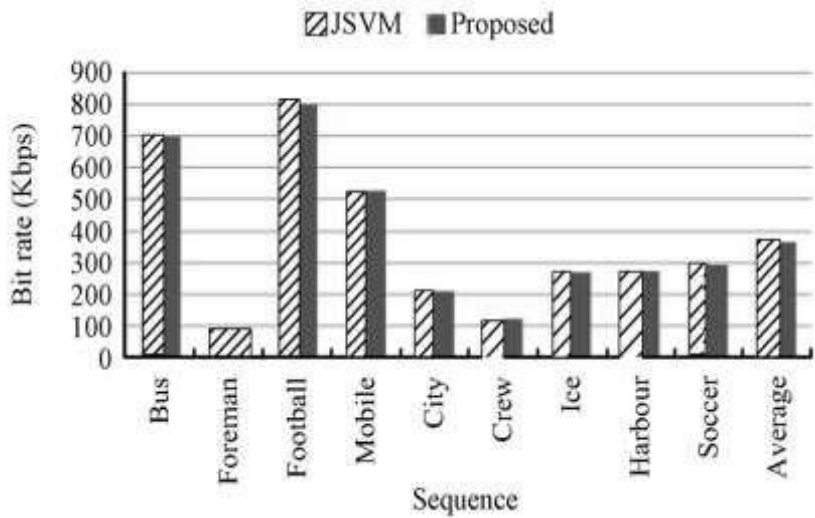

Fig.3 Comparison of average bit rate between JSVM and the proposed Algorithm

Fig.4 shows the comparison in terms of average measures between our proposed algorithm and JSVM. From the observations in three dimensions, the complexity of computation in deciding the best mode by searching all the modes is reduced in our algorithm, although our algorithm achieves higher coding efficiency by attaining the best reduction in encoding time with minimal PSNR loss.

The maximum average encoding time is achieved for Foreman, Bus, Soccer, Mobile and Foreman sequence with minimal loss in PSNR. And there was a little increase in bit rate when compared to the other chosen sequences. This is because the sequences with large motion have large change in spatial and temporal layers, in which the MB modes chosen here are based on mode correlations between the base layer and enhancement layer.

The rate distortion curves for all the 9 sequences are shown in Figs.4–13. Compared to JSVM, it is found that these sequences will not have much difference in bit rate as well as PSNR. This proves that our algorithm can be used to encode the sequences with lesser encoding time and almost the same PSNR and bit rate better and faster.

Table 1 Comparative measure between proposed algorithm and JSVM

| Sequence | Average time (SEC) | | Δ time (%) | Average Y-PSNR (dB) | | Δ PSNR (dB) | Average bit rate (Kbps) | | Δ Bit rate (%) |
|---|---|---|---|---|---|---|---|---|---|
| | JSVM | Proposed | | JSVM | Proposed | | JSVM | Proposed | |
| Bus | 69.75 | 37.13 | 46.77 | 37.34 | 37.28 | −0.05 | 698.73 | 696.60 | 0.30 |
| Foreman | 67.74 | 32.26 | 52.37 | 32.74 | 32.70 | −0.04 | 83.64 | 83.07 | 0.68 |
| Football | 149.78 | 84.94 | 43.29 | 37.30 | 37.30 | 0.00 | 800.11 | 801.24 | −0.14 |
| Moblle | 44.69 | 24.23 | 45.80 | 33.91 | 33.85 | −0.05 | 526.23 | 527.39 | −0.22 |
| Ctiy | 46.71 | 30.89 | 33.87 | 33.72 | 33.73 | 0.00 | 206.30 | 206.43 | −0.07 |
| Crew | 44.98 | 27.64 | 38.54 | 36.00 | 36.00 | 0.00 | 117.46 | 117.66 | −0.17 |
| Ice | 45.21 | 27.38 | 39.44 | 41.22 | 41.21 | −0.01 | 267.23 | 267.08 | 0.05 |
| Harbour | 40.12 | 33.01 | 17.73 | 31.15 | 31.15 | 0.00 | 269.39 | 269.37 | 0.01 |
| Soccer | 42.88 | 23.23 | 45.84 | 25.84 | 25.82 | −0.02 | 293.84 | 292.44 | 0.47 |
| Average | 61.32 | 35.63 | 41.89 | 34.36 | 34.34 | −0.02 | 362.55 | 362.37 | 0.05 |

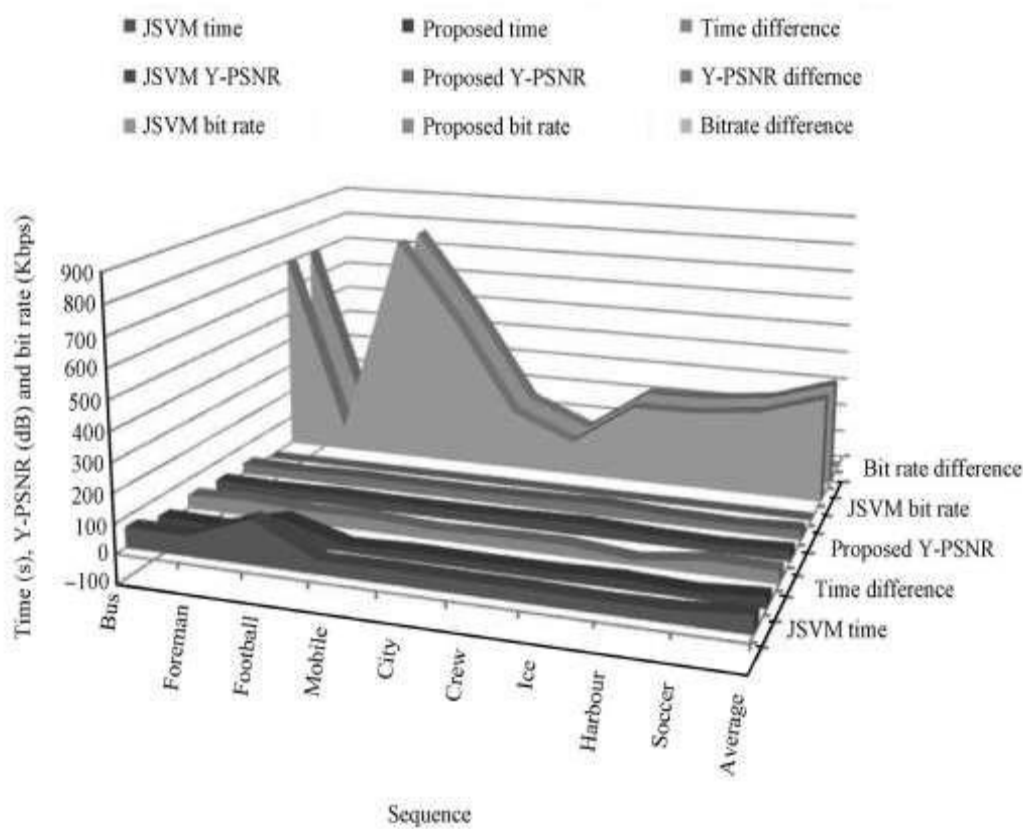

Fig.4 Comparison of average measure with JSVM and the proposed algorithm

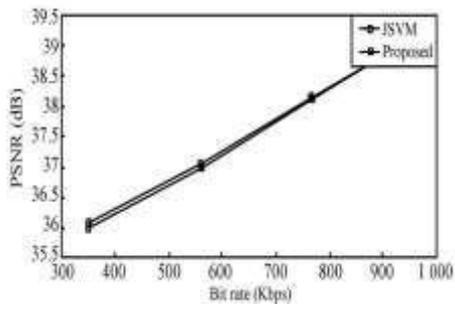
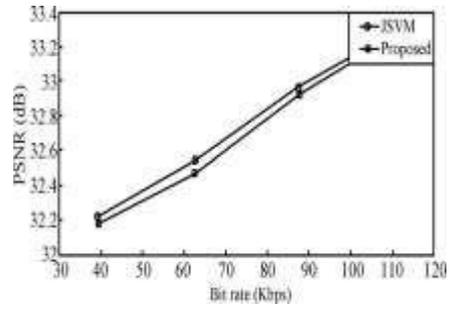

Fig.5 RD curve for Bus sequence

Fig.6 RD curve for Foreman sequence

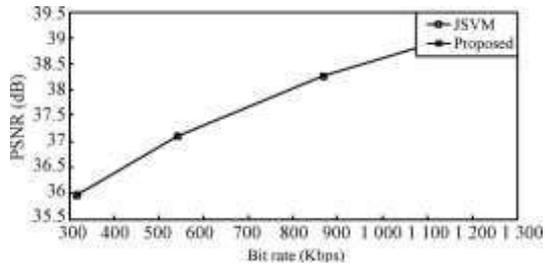
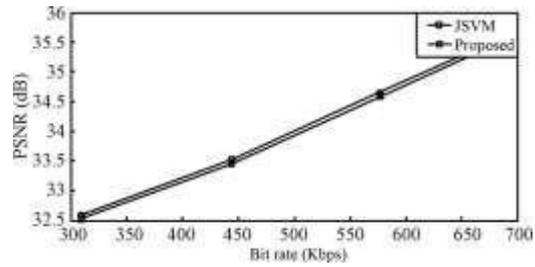

Fig.7 RD curve for Football sequence

Fig.8 RD curve for Mobile sequence

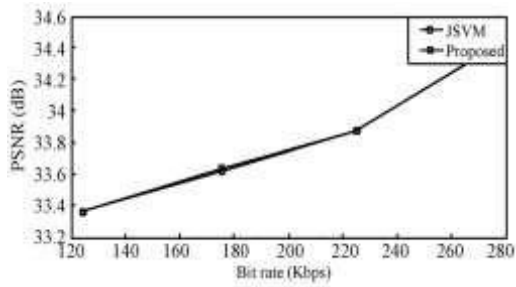
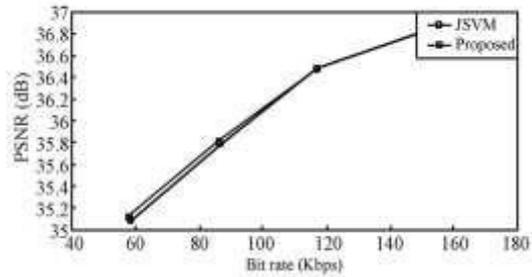

Fig.9 RD curve for City sequence

Fig. 10 RD curve for Crew sequence

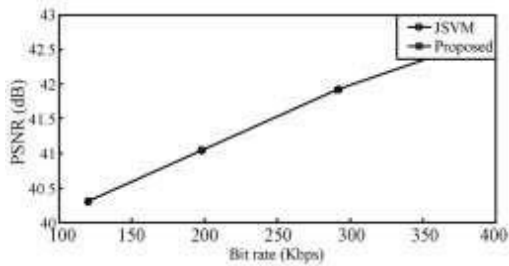
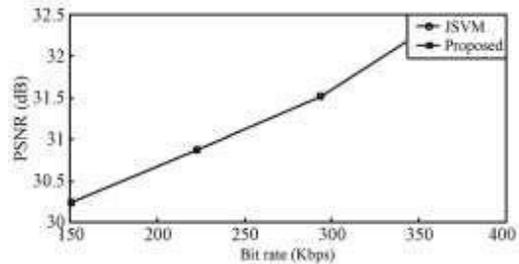

Fig. 11 RD curve for Ice sequence

Fig.12 RD curve for Harbour sequence

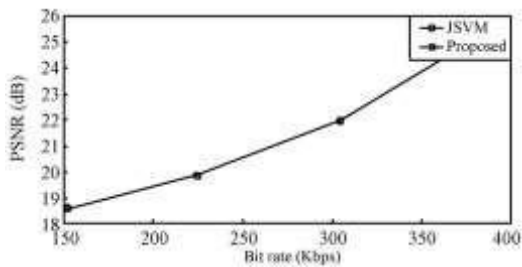

Fig.13 RD curve for Soccer sequence

The minimum average encoding time is achieved for Harbour, City, Crew and Ice sequences with improvement in PSNR as well as reduction in bit rate. It is due to small change in motion within the chosen sequences. These sequences will have small changes in the spatial and temporal level. As for our algorithm, less MVD will lead to probability model for mode selection. In our algorithm, maximum encoding time for fast motion sequences in spatial and temporal level results in minimal quality degradation and increase in bit rate. Minimum encoding time for slow motion sequences in spatial and temporal level results in maximum quality degradation and reduction in bit rate.

**5 Conclusions and future work**

A novel algorithm is proposed to decide the modes faster than JSVM by using the correlation of modes and a probability model. A desired mode list is created based on the probability model for base layer. Enhancement layer mode selection is decided by the correlation of modes among MB and current from MB. Our algorithm is implemented in JSVM 9.19.15 reference software and the performance is evaluated based on time, PSNR and bit rate. The results show 41.89% improvement in encoding time with minimal loss of 0.02dB in PSNR and 0.05% increase in bit rate. The tradeoff of maximum encoding time with minimal loss in PSNR and increase in bit rate for fast motion sequences in our algorithm may be considered as future work.

**References**

[1] J. Reichel, H. Schwarz, M. Wien. ScalFig.11 RD curve for Ice sequence able video coding, [Online], Available: http://ip.hhi.de/imagecom G1/savce/downloads/H.264MPEG4-AVCVersion8-FinalDraft.pdf, Nice, France, 2005.

[2] H. Schwarz, D. Marpe, T. Wiegand. Overview of the scalable H. 264/MPEG-4 AVC extension. In Proceedings of the International Conference on Image Processing, IEEE, Atlanta, GA, USA, pp.161-164, 2006.

[3] H. Schwarz, D. Marpe, T. Wiegand. Overview of the scalable video coding extension of the H. 264/AVC standard. IEEE Transactions on Circuits and Systems for Video Technology, vol.17, no.9, pp.1103–1120, 2007.

[4] M. Wien, H. Schwarz, T. Oelbaum. Performance analysis of SVC. IEEE Transactions on Circuits and Systems for Video Technology, vol.17, no.9, pp.1194–1203, 2007.

[5] H. Li, Z. G. Li, C. Y. Wen, Fast mode decision algorithm for inter-frame coding in fully scalable video coding. IEEE Transactions on Circuits and Systems for Video Technology, vol.16, no.7, pp.889–895, 2006.

[6] S. Lim, J. Yang, B. Jeon. Fast coding mode decision for scalable video coding. In Proceedings of the International Conference on Advanced Communication Technology, IEEE, Gangwon-Do, Korea, vol.3, pp.1897–1900, 2008.

[7] B. Lee, M. Kim, S. Hahm, C. Park, K. Park. A fast mode selection scheme in inter-layer prediction of H. 264 scalable extension coding. In Proceedings of the International Symposium on Broadband Multimedia Systems and Broadcasting, IEEE, Las Vegas, NV, USA, pp.1–5, 2008.

[8] L. Balaji, K. K. Thyagharajan. An adaptive rate control scheme for H. 264 scalable video coding. In Proceedings of the International Conference on Green Computing, Communication and Conservation of Energy, IEEE, Chennai, India, pp.40–44, 2006.

[9] T. Song, K. Takei, T. Shimamoto. Low complexity algorithm for inter-layer prediction of H.264/SVC. International Journal of Innovative Computing, Information and Control, vol.8, no.5(A), pp.3103–3113, 2012.

[10] T. S. Zhao, H. L. Wang, S. Kwong, S. D. Hu. Probability based coding mode prediction for H. 264/AVC. In Proceedings of the International Conference on Image Processing, IEEE, Hong Kong, China, pp.26–29, 2010.

[11] T. Hamamoto, T. Song, T. Katayama, T. Shimamoto. Complexity reduction algorithm for hierarchical B-picture of H.264/SVC. International Journal of Innovative Computing, Information and Control, vol.7, no.1, pp.445–457, 2011.

[12] J. Vieron, M. Wien, H. Schwarz. JSVM Reference Software, [Online], Available: http://www.hhi.fraunhofer.de/de/kompetenzfelder/imageprocessing/researchgroups/image-video-coding/svc-extension-of-h264avc/jsvm-reference-software.html, Geneva, 2014.